\documentclass[aps,pra,floatfix,amsmath,amssymb,showpacs,showkeys,twocolumn,10pt]{revtex4-1}
\usepackage[caption=false]{subfig}
\usepackage{graphicx,bm,color}
\usepackage{amsfonts}
\begin{document}

\frenchspacing

\title{Generating a perfect quantum optical vortex}

\begin{abstract}
In this article we introduce a novel quantum state, the perfect quantum optical vortex state which exhibits a highly localised distribution along a ring in the quadrature space. We examine its nonclassical properties using the Wigner function and the negativity volume. Such a quantum state can be a useful resource for quantum information processing and communication.
\end{abstract}

\author{Anindya Banerji$^{1,2}$}\email[Corresponding author: ]{abanerji09@gmail.com}
\author{Ravindra Pratap Singh$^{3}$}\email{rpsingh@prl.res.in}
\author{Dhruba Banerjee$^{1}$}\email{dhruba.iacs@gmail.com}
\author{Abir Bandyopadhyay$^{2}$}\email{abir@hetc.ac.in}
\affiliation{$^{1}$Department of Physics, Jadavpur University, Kolkata 700032, India}
\affiliation{$^{2}$Hooghly Engineering and Technology College, Hooghly 712103, India}
\affiliation{$^{3}$Physical Research Laboratory, Ahmedabad 380009, India}

\date{\today}


\maketitle

\section{Introduction}

Implementation of quantum information protocols using the various tools at the disposal of quantum optics has always resulted in intriguing study. This field of research has seen intense activity since quite some time and became even more interesting ever since the seminal KLM proposal \cite{knill2001scheme}. One of the most potent tools used by researchers in this field for the generation of entangled pairs of particles, namely photons, is the spontaneous parametric down conversion (SPDC) using non linear beta barium borate (BBO) crystals. In this unique process, a single photon splits into two photons which are entangled in multiple degrees of freedom (DOF). This process also leads to the formation of hyperentangled pairs of photons. In hyperentanglement, the pair of particles is entangled in all the available DOF \cite{Kwiat}. The uniqueness of SPDC lies in the fact, that the photon pair can be treated to be entangled in any particular degree of freedom as required by the application protocol irrespective of the other DOF. For example, the output from type II SPDC can be described by the following expression
\begin{equation}
\label{polarization}
\vert\Psi\rangle_{SPDC}=\frac{1}{\sqrt{2}}\left(\vert H_1\rangle\vert V_2\rangle+\vert V_1\rangle\vert H_2\rangle\right)
\end{equation}
\noindent where $ {H, V} $ denote the polarization state of the photons arriving in diametrically opposite points of the SPDC ring. It means that if the first photon is found to be horizontally polarized, the second photon would be vertically polarized and vice versa. The same output, can be also written as
\begin{equation}
\label{oam}
\vert\Phi\rangle=\sum_lC_{l_1,l_2}\vert l\rangle\vert -l\rangle
\end{equation}
\noindent where \emph{l} denotes the OAM value and the state is seen to be entangled in OAM \cite{Mair}. Since OAM values can range from $-\infty$ to $+\infty$, such a state can exhibit very high degree of entanglement. Moreoever, systems like this can be used in the implementation of complete Bell state analysis \cite{walborn2003hyperentanglement, sheng2009deterministic} with 100$\%$ efficiency using only linear elements \cite{Kwiat1998LinearElements}. Besides, hyperentangled systems exhibit significant advantages in quantum communication protocols like superdense coding \cite{FromKwiat1} and quantum crytography \cite{FromKwiat2}.\\
In general, photon pairs produced in SPDC output belong to the larger class of Gaussian states and has been subjected to detailed study and scrutiny over the years in the context of quantum information \cite{gaussianstates}. Lately though, there has been increasing focus on a new class of states being generated from SPDC photons called the quantum vortex states \cite{GSA97, jbvortex, abir, GSANJP, JPhysA, OptComm, ArXiv}. These are non-Gaussian quantum states exhibiting phase singularity or topological defect. A vortex state of topological index \emph{l} carries with it an orbital angular momentum $l\hbar$. It has also been reported that higher the order, higher is the entanglement \cite{JPhysA}. Although \emph{l} can range from $0$ to $\infty$, neglecting the handedness, in practice, it is observed, higher the order, larger is the spread of the vortex core. A way out of this is the generation of perfect quantum optical vortex states (PQOVS). The spread of the vortex core do not change with the order of OAM in such PQOVS. In this article we report the generation of the PQOVS and study its various properties through the use of quadrature distribution and Wigner function.\\
In order to generate the PQOVS, we begin with one of the simplest non - Gaussian state, the pair coherent states. Pair coherent states, introduced by Agarwal, are coherent states with a fixed difference in the number of photons between the modes. The difference in the number of photons fixes the order or topological charge of the state. These states are Bessel - Gaussian in nature. We perform a Fourier transform on pair coherent states to generate the perfect vortex.\\
The article is organized as follows. In section \ref{sec:QuantumPerfectVortex} we present a detailed mathematical study on the generation of PQOVS. In section \ref{sec:WignerFunction}, we derive the Wigner function associated with PQOVS explicitly and present our interpretations of the results obtained therein. We conclude the article in section \ref{sec:Conclusion} with a brief review of the results and directions for future work.

\section{Generating a perfect quantum vortex}
\label{sec:QuantumPerfectVortex}
\begin{figure*}
\centering
\subfloat[q=1]{
\includegraphics[scale=0.2]{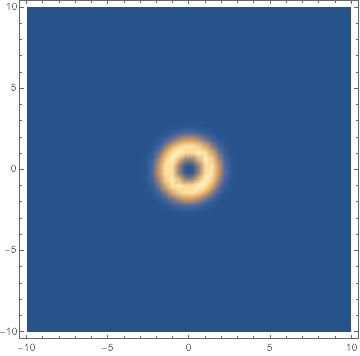}
\label{fig:BGcontour1}}
\qquad
\subfloat[q=10]{
\includegraphics[scale=0.2]{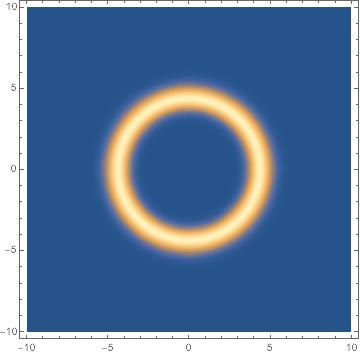}
\label{fig:BGcontour10}}
\qquad
\subfloat[q=15]{
\includegraphics[scale=0.2]{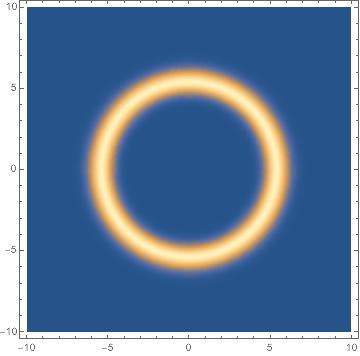}
\label{fig:BGcontour15}}
\qquad
\subfloat[q=20]{
\includegraphics[scale=0.2]{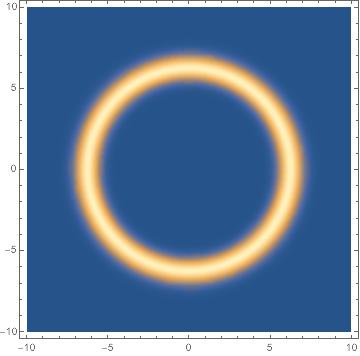}
\label{fig:BGcontour20}}
\qquad\\
\subfloat[q=1]{
\includegraphics[scale=0.2]{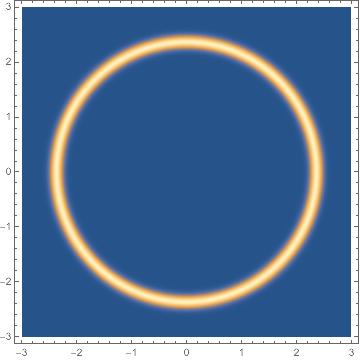}
\label{fig:perfectcontour1}}
\qquad
\subfloat[q=10]{
\includegraphics[scale=0.2]{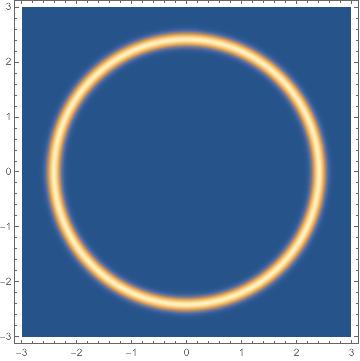}
\label{fig:perfectcontour10}}
\qquad
\subfloat[q=15]{
\includegraphics[scale=0.2]{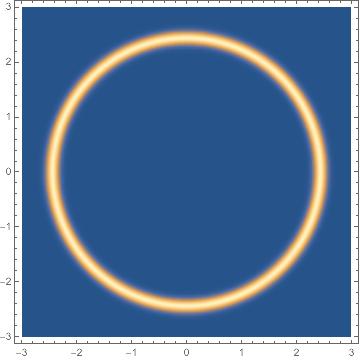}
\label{fig:perfectcontour15}}
\qquad
\subfloat[q=20]{
\includegraphics[scale=0.2]{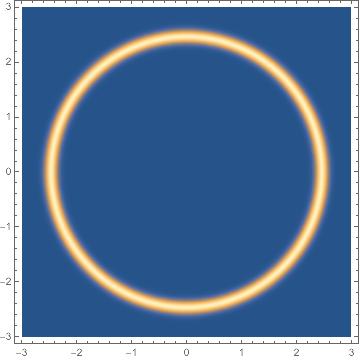}
\label{fig:perfectcontour20}}
\caption{(Color online) Perfect quantum optical vortex states for various orders (e - h) in comparison with the corresponding BG vortex states. \emph{q} represents the topological charge or order of the vortex state.}
\label{fig:quadratures}
\end{figure*}
A perfect optical vortex, in theory, is best represented by the equation
\begin{equation}
\label{classicalperfectvortex}
E\left(r,\theta\right)=\frac{\text{i}^{l-1}}{k_r}\delta\left(r-r_c\right)\exp\left(\text{i}l\theta\right)
\end{equation}
\noindent which represents an ideal perfect optical vortex beam with the central dark core having a radius $r_c$. Such a vortex beam can be generated by Fourier transforming an ideal Bessel beam using a simple lens \cite{Vaity}. But, as noted by the authors in the same work, generation of ideal Bessel beam is not possible experimentally. A way out of this is by using a Bessel - Gauss beam in place of ideal Bessel beam of the form
\begin{equation}
\label{BGbeam}
E(\rho,\phi)=\text{J}_l(k_r,\rho)\exp(\text{i}l\phi)\exp\left(-\frac{\rho^2}{\sigma^2}\right)
\end{equation}
\noindent where $J_l(k_r,\rho)$ is the \emph{l}th order Bessel function of the first kind. $k_r$ is the longitudinal wave vector and the second exponential term on the right represents the Gaussian wavepacket that confines the Bessel beam. In order to generate a perfect quantum optical vortex, one should begin with a state that reduces to the form of Eq. \ref{classicalperfectvortex} in the quadrature distribution. To that end, we begin with one of the simplest non - Gaussian quantum states, often referred to as pair coherent states \cite{AgarwalPaircoherent} in the literature. In the coherent state representation, these states can be written as
\begin{equation}
\label{paircoherentstates}
\vert\zeta,q\rangle=\frac{\text{e}^q}{2\pi\sqrt{\zeta^q\text{I}_0\left(2\zeta\right)}}\int^{2\pi}_0\left(\sqrt{\zeta}\text{e}^{\text{i}\theta}\right)^{-q}\vert\sqrt{\zeta}\text{e}^{\text{i}\theta}\rangle\vert\sqrt{\zeta}\text{e}^{-\text{i}\theta}\rangle\text{d}\theta
\end{equation}
\noindent where $\text{I}_0$ represents the modified Bessel function of the first kind of order \emph{0}. $\zeta$ is the complex amplitude. The meaning of \emph{q} becomes clear when Eq. \ref{paircoherentstates} is written in the Fock basis as follows
\begin{equation}
\label{paircoherentstateinfockbasis}
\vert\zeta,q\rangle_{\text{F}}=\frac{1}{\sqrt{\zeta^q\text{I}_0\left(2\zeta\right)}}\sum^{\infty}_{n=0}\frac{\zeta^n}{\sqrt{n!\left(n+q\right)!}}\vert n+q,n\rangle
\end{equation}
As is evident from the above equation, \emph{q} is the number of photons added to one of the modes of a two mode state. Or in other words, it is the difference in the number of photons present in each mode. Under conditional parametric conversion process, a pair coherent state evolves to a vortex state as observed in \cite{BGV}. The Hamiltonian describing such a process has the following form
\begin{equation}
\label{Hamiltonian}
H=\omega_aa^{\dagger}a+\omega_bb^{\dagger}b+\kappa\left(a^{\dagger}b\text{e}^{\text{i}\eta t}+ab^{\dagger}\text{e}^{-\text{i}\eta t}\right)
\end{equation}
where $\eta$ is the difference in frequencies and $\kappa$ is the coupling constant. Under this Hamiltonian, Eq. \ref{paircoherentstates} evolves into
\begin{equation}
\label{BGvortex}
\vert\psi\rangle=\mathcal{N}\int_0^{2\pi}\text{e}^{\text{i}q\theta}\vert\zeta\cos\theta\rangle\vert\zeta\sin\theta\rangle\text{d}\theta
\end{equation}
\noindent which has a vortex structure in the quadrature distribution. Eq. \ref{BGvortex} represents a BG vortex state \cite{BGV} of topological charge \emph{q}. It carries with it orbital angular momentum $q\hbar$. $\mathcal{N}$ is the normalization factor and written as
\begin{equation}
\label{normalization}
\mathcal{N}=\frac{1}{\sqrt{4\pi^2\text{e}^{-\alpha^2}\text{I}_q\left(\alpha^2\right)}}
\end{equation}
\noindent where $\text{I}_q$ is the \emph{q}th order modified Bessel function of first kind. In the cylindrical coordinates, this state has the form
\begin{equation}
\label{cylindricalBGvortex}
\psi(\rho,\phi)=2\text{i}^q\sqrt{\pi}\mathcal{N}\exp\left(-\frac{\rho^2}{2}\right)\text{J}_q\left(\sqrt{2}\alpha \rho\right)\exp\left(\text{i}q\phi\right)
\end{equation}
\noindent where $\alpha=\vert\zeta\vert$ and $\text{J}_q$ is the \emph{q}th order Bessel function of the first kind.
%beginning figure 
\begin{figure}
\centering
\subfloat[q=1]{
\includegraphics[scale=0.2]{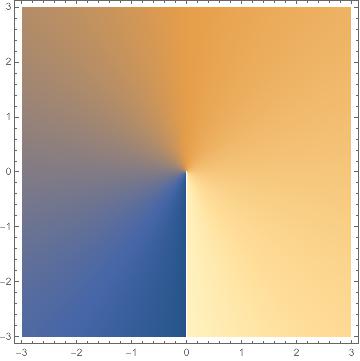}
\label{fig:perfectphase1}}
\qquad
\subfloat[q=10]{
\includegraphics[scale=0.2]{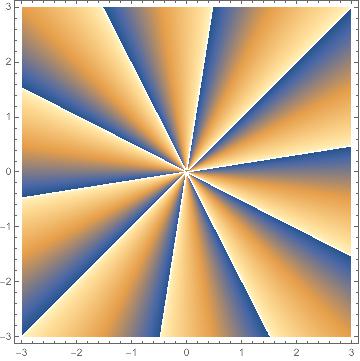}
\label{fig:perfectphase10}}
\qquad
\subfloat[q=15]{
\includegraphics[scale=0.2]{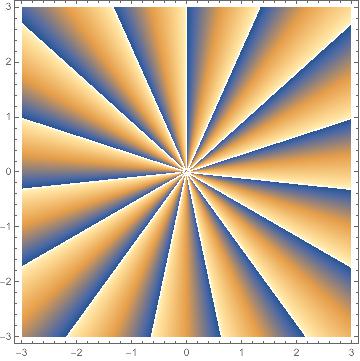}
\label{fig:perfectphase15}}
\qquad
\subfloat[q=20]{
\includegraphics[scale=0.2]{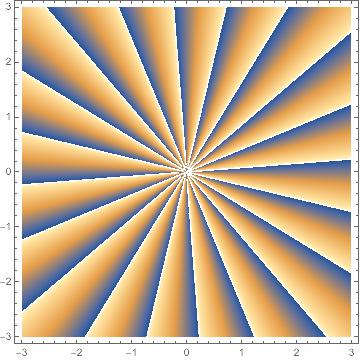}
\label{fig:perfectphase20}}
\caption{(Color online) Phase of the generated PQOVS for various orders.}
\label{fig:phase}
\end{figure}
Any arbitrary optical beam can be Fourier transformed by passing it through a simple lens. The transformation equation corresponding to such a process, for an arbitrary beam $E\left(\rho,\phi\right)$ can be expressed as \cite{introtofourieroptics}
\begin{eqnarray}
\label{opticalfouriertransform}
E\left(r,\theta\right)=\frac{k}{2\pi\text{i}f}\int_0^{\infty}\int_0^{2\pi}E\left(\rho,\phi\right)\rho\text{d}\rho\text{d}\phi\nonumber\\
\times\exp\left(-\frac{\text{i}k}{f}\rho r\cos \left(\theta-\phi\right)\right)
\end{eqnarray}
\noindent in cylindrical coordinate system. Here \emph{f} is the focal length of the lens used and \emph{k} is the wave vector. Such a transformation on an ideal Bessel beam gives rise to Eq. \ref{classicalperfectvortex}. Substituting $E\left(\rho,\phi\right)$ with Eq. \ref{cylindricalBGvortex}, the Fourier transform of the BG vortex is
\begin{eqnarray}
\label{quadratureperfectBGvortex}
\psi\left(r,\theta\right)&=&\frac{k\text{i}^{q-1}}{\sqrt{\pi}f}\mathcal{N}\int_0^{\infty}\int_0^{2\pi}\exp\left(-\frac{\rho^2}{2}\right)\text{J}_q\left(\sqrt{2}\alpha \rho\right)\nonumber\\
 &\times &
\exp\left(\text{i}q\phi\right)\rho\text{d}\rho\text{d}\phi\exp\left(-\frac{\text{i}k}{f}\rho r\cos \left(\theta-\phi\right)\right)
\end{eqnarray}
Evaluating the integration in Eq. \ref{quadratureperfectBGvortex} using standard integral techniques and applying the properties of the Bessel function, we arrive at the following form
\begin{eqnarray}
\label{quadratureperfectBGvortexfinal}
\psi\left(r,\theta\right)&=&\mathcal{N}\frac{2\sqrt{2}\pi}{\sigma}\text{i}^{2q-1}\exp\left(\text{i}q\theta\right)
\nonumber\\ &\times &\exp\left[-\frac{r_c^2+r^2}{\sigma^2}\right]\text{I}_q\left(\frac{2r_cr}{\sigma^2}\right)
\end{eqnarray}
\noindent where $r^2=x^2+y^2$ and $\theta=\tan^{-1}\left(y/x\right)$ are polar coordinates and $\sigma=\sqrt{2}f/k$. $r_c=\sqrt{2}\alpha f/k$ is the radius of the central dark core. It is dependent on the amplitude of the coherent state, the focal length of the lens used in the Fourier transformation and the wave vector of the incident beam. This equation resembles the one obtained in \cite{Vaity} with one vital difference. The core radius $r_c$ is a function of the coherent state amplitude and resembles the role played by the axicon parameter in \cite{Vaity}. On further comparison, we note that $\sigma$ can be interpreted as the Gaussian beam waist at the focus. Eq. \ref{quadratureperfectBGvortexfinal} represents the quadrature distribution associated with the perfect quantum optical vortex state.\\
We study the perfect quantum optical vortex (PQOVS) state in Fig. \ref{fig:quadratures}. We have used $\lambda=810 nm$ and focal length of the lens $f=70 cm$ for all the figures. In case of $\alpha$, it was observed that a higher value resulted in more stability of the core radius for higher order PQOVS. An advantage of this scheme over the classical scheme is that in the later the axicon parameter had to be modified to keep the core radius fixed with changing order of the resultant state. That would involve modification of the experimental setup. But in our scheme, only modifying the coherent state amplitude imparts far greater stability to the core radius even for very high orders of the PQOVS. In our study, we have used $\alpha=15$. We observe that the radius of the dark core remains almost constant for different orders of the vortex or topological charge. In contrast, the central core radius of a standard BG vortex varies quite substantially with changing values of \emph{q}. Slight variation in the core radius is also observed for the PQOVS and can be attributed to the rate of change of $\text{I}_q$ with \emph{q} as noted in \cite{Vaity}. But this change is almost negligible when compared to the change observed in the case of a standard BG vortex. The order or the topological charge of the generated PQOVS can be confirmed by studying the phase of the associated quadrature distribution of Eq. \ref{quadratureperfectBGvortexfinal} which we study in Fig. \ref{fig:phase}. We observe a single discontinuity for $q=1$ and ten discontinuities for $q=10$, fifteen for $q=15$ and twenty for $q=20$ which corresponds to the order of the PQOVS.

\section{Wigner Function of the PQOVS}
\label{sec:WignerFunction}
\begin{figure}
\centering
\subfloat[$W(x, y)_{p_y=0}^{p_x=0}$]{
\includegraphics[scale=0.25]{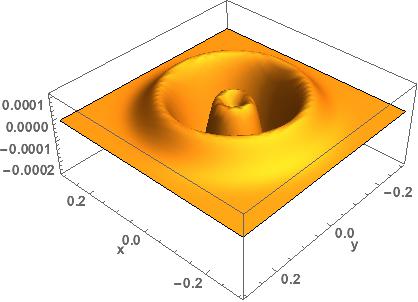}
\label{fig:wignerxy}}
\qquad
\subfloat[$W(x, p_x)_{y=0}^{p_y=0}$]{
\includegraphics[scale=0.25]{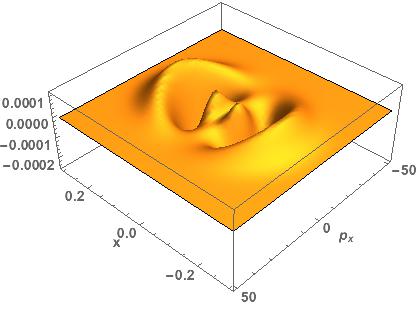}
\label{fig:wignerxpx}}
\qquad\\
\subfloat[$W(x, p_y)_{y=0}^{p_x=0}$]{
\includegraphics[scale=0.25]{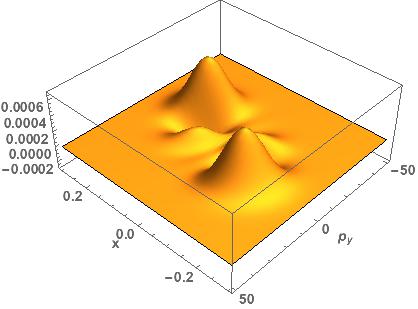}
\label{fig:wignerxpy}}
\qquad
\subfloat[$W(y, p_y)_{x=0}^{p_x=0}$]{
\includegraphics[scale=0.25]{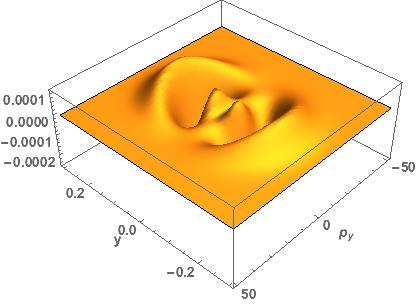}
\label{fig:wignerypy}}
\qquad\\
\subfloat[$W(y, p_x)_{x=0}^{p_y=0}$]{
\includegraphics[scale=0.25]{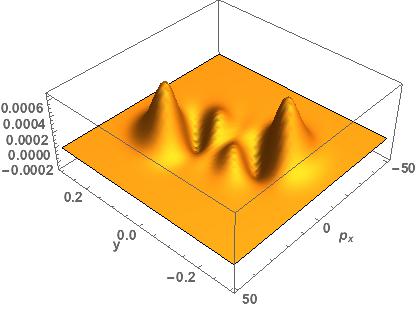}
\label{fig:wignerypx}}
\qquad
\subfloat[$W(p_x, p_y)_{x=0}^{y=0}$]{
\includegraphics[scale=0.25]{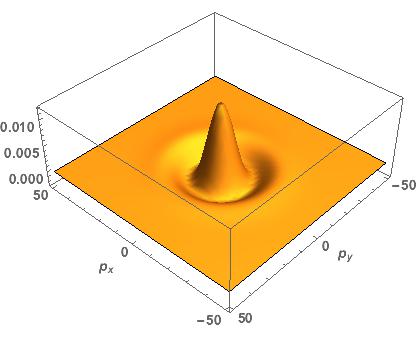}
\label{fig:wignerpxpy}}
\caption{(Color online) Wigner function $W$ associated with the PQOVS for $q=2$.}
\label{fig:Wigner}
\end{figure}
In this section, we derive the Wigner function \cite{wigner1932quantum} associated with a PQOVS. The Wigner function is a quasi probability distribution. It differs from a classical probability distribution due to the fact that it takes negative values for which there are no classical analogues. It provides direct evidence of non locality for EPR systems \cite{banaszek1998nonlocality}. It can also be interpreted as the Fourier transform of the two point correlation function. The marginals of the Wigner function should be interpreted as partial trace operations. The reduced Wigner function, that is the case when any one (two) quadrature(s) is (are) taken as the argument while the other one (two) is (are) treated as parameter(s) for a single (two) mode state should be interpreted as a conditional probability distribution where the outcome of one is conditioned upon the outcome of the other. It is in this respect, the negativity of the Wigner function, which has been interpreted as a signature of non classicality \cite{kenfack2004negativity}, assumes greater importance. Here we would like to quote Feynman ``If a physical theory for calculating probabilities yields a negative probability for a given situation under certain assumed conditions, we need not conclude the theory is incorrect. Two other possibilities of interpretation exist. One is that the conditions (for example, initial conditions) may not be capable of being realised in the physical world. The other possibility is that the situation for which the probability appears to be negative is not one that can be verified directly. A combination of these two, limitation of verifiability and freedom in initial conditions, may also be a solution to the apparent difficulty" \cite{FeynmanNegativeProbability}. In this context, studying the Wigner function associated with our proposed state seems all the more interesting. The Wigner function for any arbitrary state $\psi(\vec{r})$ of two modes is defined as
\begin{eqnarray}
\label{wignerdefinition}
W &=& \frac{1}{\left(2\pi\right)^2}\int_{-\infty}^{\infty}\psi\left(\vec{r}+\vec{R/2}\right)\psi^*\left(\vec{r}-\vec{R/2}\right)\nonumber\\
&\times &\exp\lbrace\text{i}\left(R_1p_1+R_2p_2\right)\rbrace\text{d}R_1\text{d}R_2
\end{eqnarray}
\noindent where $\vec{r}=x+\text{i}y$ and $\vec{R}=R_1+\text{i}R_2$. Rewriting Eq. \ref{quadratureperfectBGvortexfinal} in terms of coordinates \emph{x} and \emph{y}, where \emph{x} stands for mode \emph{a} and \emph{y} stands for mode \emph{b}, we get
\begin{equation}
\label{cartesianPBGV}
\psi\left(x,y\right)=A\text{i}^{2q-1}\left[x+\text{i}y\right]^q\exp\left(-\frac{x^2+y^2}{\sigma^2}\right)\text{I}_q\left(B\sqrt{x^2+y^2}\right)
\end{equation}
\noindent where $A=\frac{\mathcal{N}2\sqrt{2}\pi}{\sigma}\exp\left(-r_c^2/\sigma^2\right)$ and $B=2r_c/\sigma^2$. Substituting Eq. \ref{cartesianPBGV} in Eq. \ref{wignerdefinition} and evaluating the integral using standard techniques \cite{Gradshteyn}, we can write the Wigner function associated with PQOVS as follows
\begin{eqnarray}
\label{finalwigner}
W=\frac{\vert A\vert^2}{4\pi^2}\text{e}^{-2\vert r\vert^2/\sigma^2}\text{e}^{-\vert p\vert^2\sigma^2/2}(-1)^q\sum_{k=0}^q k!\nonumber\\
2^{1-q}\pi\sigma^{2q+2}\text{I}_q\left(B\vert r\vert^2\right)\nonumber\\
\times\text{L}_q\left[\frac{4\vert r\vert^2 +\vert p\vert^2 \sigma^4 +4\left(p_xy-p_yx\right)\sigma^2}{2\sigma^2}\right]
\end{eqnarray}
\noindent where $\vert p\vert^2=p_x^2+p_y^2$ and $p_x$ and $p_y$ are conjugate momenta of the two modes. $\text{L}_q$ represents the Laguerre polynomial of order q. We observe that the final form of the Wigner function consists of a Gaussian factor composed of quadratures of both the modes, a modified Bessel function of order \emph{q} only in \emph{r} and a Laguerre polynomial, also of order \emph{q} which takes a combination of the components of $\vec{r}$ and $\vec{p}$ as arguments. We study the properties of Eq. \ref{finalwigner} in more detail in Fig. \ref{fig:Wigner}. Since \emph{W} is \emph{4}-dimensional in structure, it is impossible to reproduce the full form of \emph{W} on paper. Hence we study the nature of the Wigner function for six possible combinations of the quadratures taking any two as parameters and the other two as variables at a certain instant in Fig. \ref{fig:Wigner}. It is observed that all the six combinations, i.e. $W\left(x,y\right)$, $W\left(x,p_x\right)$, $W\left(x,p_y\right)$, $W\left(y,p_y\right)$, $W\left(y,p_x\right)$ and $W\left(p_x,p_y\right)$ show negative regions. This negativity has often been interpreted as an indicator of quantum interference. The Wigner function corresponding to the quadratures \emph{x} and \emph{y}, Fig. \ref{fig:wignerxy} and $p_x$ and $p_y$, Fig. \ref{fig:wignerpxpy}, exhibit similar nature. It consists of concentric rings of diminishing intensity with the intensity diminishing while moving outwards from the centre. Similar quadratures of the two modes (spatial quadratures in Fig. \ref{fig:wignerxy} and momentum quadratures in Fig. \ref{fig:wignerpxpy}) oscillate in orthogonal directions to give rise to the concentric rings which are bounded by the Gaussian function and hence the rings decrease in magnitude. In case of $W\left(x,p_x\right)$, both quadratures of a mode overlap to give rise to the pattern seen in Fig. \ref{fig:wignerxpx}. A similar explanation holds for Fig. \ref{fig:wignerypy}. The interesting results appear in Fig. \ref{fig:wignerxpy} and Fig. \ref{fig:wignerypx}. A correlation is observed between the spatial quadrature of one mode and the mmomentum quadrature of the other mode. The spatial quadrature of one mode interacts with the momentum quadrature of the other mode in absence of the spatial quadrature of the later and momentum quadrature of the former to give rise to interference like structure with negative regions. These structures have often been referred to as quantum interference pattern and the associated negative volume used to study nonclassical correlations. The negative volume of the Wigner function is defined as
\begin{equation}
n(W)=\frac{1}{2}\int\int\vert W\left(x,p_y\right)\vert \text{d}x\text{d}p_y-1
\end{equation}
\begin{figure}
\centering
\includegraphics[scale=0.4]{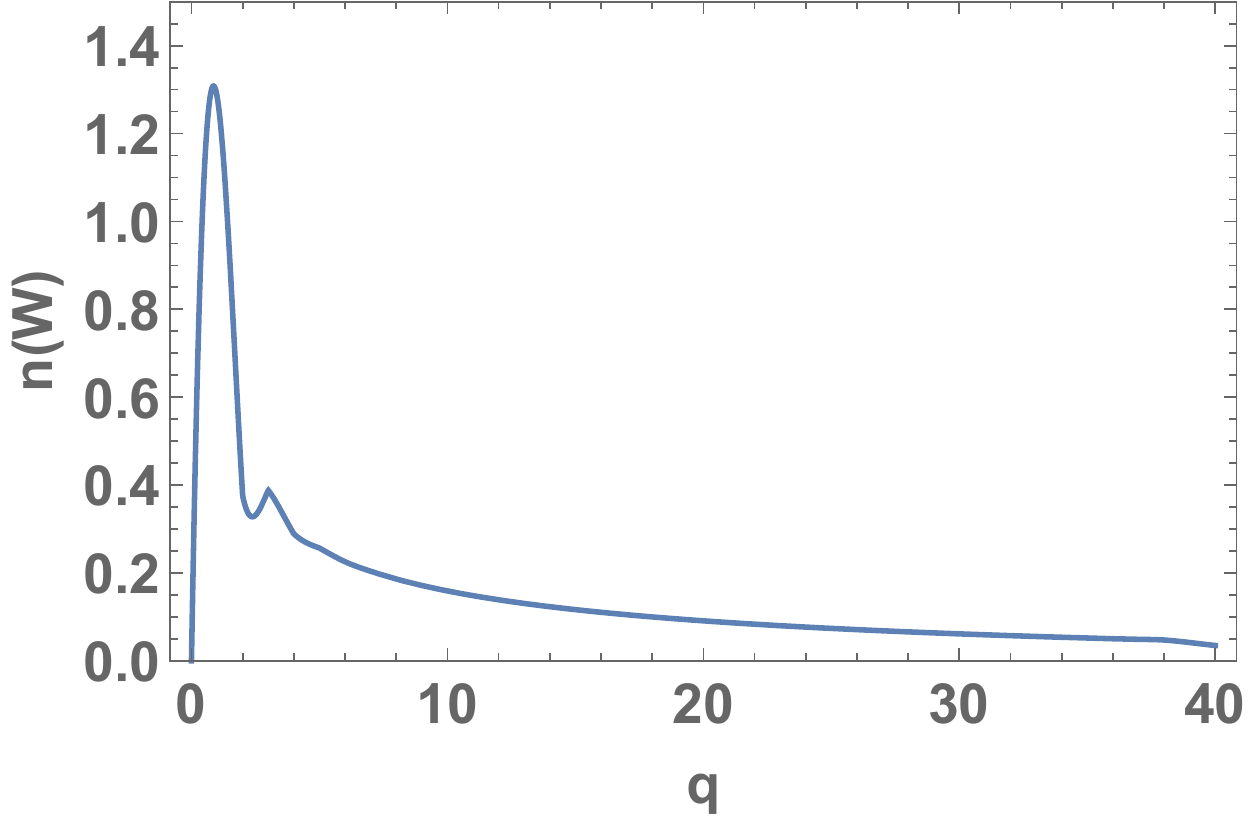}
\caption{(Color online) Variation of the negative volume of $W(x,p_y)$ with $y=0;p_x=0$ with order \emph{q}.}
\label{fig:NegativeVol}
\end{figure}
\begin{figure}
\centering
\subfloat[$W(x,p_y)_{q=0}^{y=0,p_x=0}$]{
\includegraphics[scale=0.25]{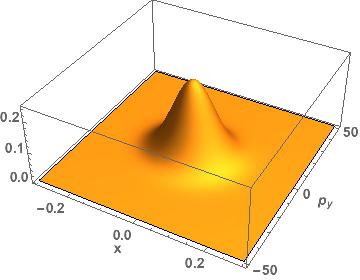}
\label{fig:Wxpyq0}}
\qquad
\subfloat[$W(x,p_y)_{q=1}^{y=0,p_x=0}$]{
\includegraphics[scale=0.25]{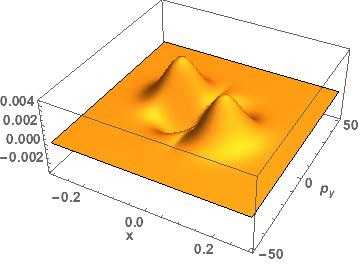}
\label{fig:Wxpyq1}}
\qquad\\
\subfloat[$W(x,p_y)_{q=2}^{y=0,p_x=0}$]{
\includegraphics[scale=0.25]{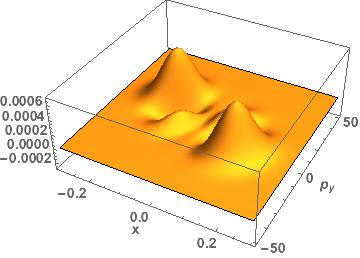}
\label{fig:Wxpyq2}}
\qquad
\subfloat[$W(x,p_y)_{q=3}^{y=0,p_x=0}$]{
\includegraphics[scale=0.25]{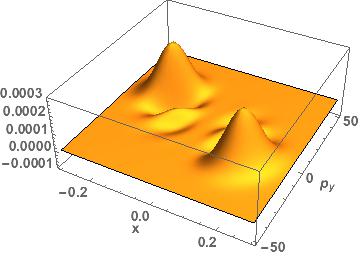}
\label{fig:Wxpyq3}}
\qquad\\
\subfloat[$W(x,p_y)_{q=4}^{y=0,p_x=0}$]{
\includegraphics[scale=0.25]{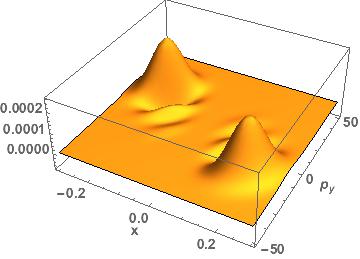}
\label{fig:Wxpyq4}}
\qquad
\subfloat[$W(x,p_y)_{q=5}^{y=0,p_x=0}$]{
\includegraphics[scale=0.25]{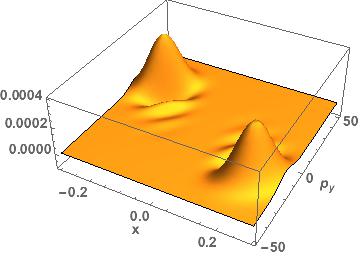}
\label{fig:Wxpyq5}}
\caption{(Color online) Wigner function $W\left(x,p_y\right)$ for different orders of the PQOVS. The subscript \emph{q} denotes the order of the vortex.}
\label{fig:QuantumInterference}
\end{figure}
We study the variation of \emph{n(W)} is Fig. \ref{fig:NegativeVol}. It is observed that the negative volume associated with $W(x,p_y)$ increases sharply with increasing order of the vortex \emph{q} before reaching a peak value of 1.287 for $q=2$ after which it decays asymtotically. This can be explained by the fact that the interference decreases with increasing value of \emph{q}. For \emph{q=0} there is only peak with no interference. The interference pattern first appears for \emph{q=1} and is the most prominent for \emph{q=2}. With further increase in \emph{q}, the interference pattern starts to separate out becoming sparse and localized in two distinct regions with the separation distance increasing with increasing order of the vortex. This is evident from Fig. \ref{fig:QuantumInterference}. The study of the negative volume can thus be used to state conclusively that the quantum interference effect decreases with increasing order of the vortex but it remains inconclusive with regards to nonclassicality.

\section{Conclusion}
\label{sec:Conclusion}
To conclude, we have introduced and theoretically studied the generation of a new non - Gaussian state with strong underlying non classical properties. We call this state perfect BG vortex state or PQOVS. We have shown that a standard BG quantum vortex evolves into a PQOVS under Fourier transformation using a simple lens. The PQOVS thus generated has a fixed core radius which can be controlled by varying the initial conditions to suit specific applications. As an advantage over recently introduced BG vortex states, the central dark core of PQOVS does not change noticeably with increasing order or topological charge of the vortex. Slight change in the radius can be offset by slighlty altering the coherent state amplitude of the initial pair coherent state. We have also calculated the Wigner function of the PQOVS. It is observed that all the six combinations of the 4-D Wigner function exhibit negative values which is a strong indication of non - classical behavior. Interesting patterns are also observed in the study of $W(x,p_y)$ and $W(y,p_x)$ which can be interpreted as a signature of quantum interference between the two modes of the PQOVS state. The negative volume of $W(x,p_y)$ is calculated for various orders of the vortex. It is observed that the negative volume peaks for $q=2$ after which it decays asymtotically with increasing order. This is because the quantum interference pattern in $W(x,p_y)$ is most pronounced for $q=2$ after which the interference pattern starts to disintegrate into two parts and appears to be localized in two distinct locations with no interaction between the two parts. The negativity of the Wigner function can therefore be used as a measure of the quantum correlation/interference between constituent modes but no conclusion can be drawn on the behavior of nonclassicality of the PQOVS with increasing order of the vortex.

\acknowledgments{This work is partially supported by DST through SERB grant no.: SR/S2/LOP - 001/2014. One of us (A. Banerji) would like to thank Physical Research Laboratory for hospitality where part of the work was done.}

\bibliographystyle{unsrt}
\bibliography{BGVortexBibliography,MyPapers}

\end{document}